\documentclass[aps,prl,twocolumn,superscriptaddress]{revtex4-1}

\usepackage{graphicx}

\usepackage{dcolumn}
\usepackage{bm}
\usepackage{natbib}
\usepackage{xcolor}
\usepackage{subcaption}
\usepackage{graphicx}
\usepackage{mwe}
\usepackage{appendix}

\makeatletter
     \renewcommand\@make@capt@title[2]{%
      \@ifx@empty\float@link{\@firstofone}{\expandafter\href\expandafter{\float@link}}%
       {\textbf{#1}}\@caption@fignum@sep#2\quad}%
     \makeatother
 
\makeatletter 
\renewcommand{\fnum@figure}{\textbf{Figure~\thefigure}}
\makeatother

\begin{document}

\title{Masses of short-lived $^{49}$Sc, $^{50}$Sc, $^{70}$As, $^{73}$Br and stable $^{196}$Hg isotopes}

\author{I. Kulikov}
\affiliation{GSI Helmholtzzentrum f\"{u}r Schwerionenforschung GmbH, Planckstra\ss e 1, 64291 Darmstadt, Germany}
\email[Corresponding author: ]{ivan.kulikov@cern.ch}

\author{A. Algora}
\affiliation{Instituto de F{\'i}sica Corpuscular, CSIC–Universitat de Valencia, E-46071 Valencia, Spain}

\author{D. Atanasov}
\affiliation{CERN, 1211 Geneva, Switzerland}

\author{P. Ascher}
\affiliation{CENBG, Gradignan, France}

\author{K. Blaum}
\affiliation{Max-Planck-Institut f\"{u}r Kernphysik, Saupfercheckweg 1, 69117 Heidelberg, Germany}

\author{R.B. Cakirli}
\affiliation{Istanbul University, Istanbul, Turkey}

\author{F. Herfurth}
\affiliation{GSI Helmholtzzentrum f\"{u}r Schwerionenforschung GmbH, Planckstra\ss e 1, 64291 Darmstadt, Germany}

\author{A. Herlert}
\affiliation{FAIR GmbH, Planckstra\ss e 1, 64291 Darmstadt, Germany}

\author{W. J. Huang}
\affiliation{Max-Planck-Institut f\"{u}r Kernphysik, Saupfercheckweg 1, 69117 Heidelberg, Germany}
\affiliation{CSNSM-IN2P3-CNRS, Universit\'{e} Paris-Sud, 91405 Orsay, France}

\author{J. Karthein}
\affiliation{CERN, 1211 Geneva, Switzerland}
\affiliation{Max-Planck-Institut f\"{u}r Kernphysik, Saupfercheckweg 1, 69117 Heidelberg, Germany}

\author{Yu. A. Litvinov}
\affiliation{GSI Helmholtzzentrum f\"{u}r Schwerionenforschung GmbH, Planckstra\ss e 1, 64291 Darmstadt, Germany}

\author{D. Lunney}
\affiliation{CSNSM-IN2P3-CNRS, Universit\'{e} Paris-Sud, 91405 Orsay, France}

\author{V. Manea}
\affiliation{KU Leuven, Instituut voor Kern- en Stralingsfysica, 3001 Leuven, Belgium}

\author{M. Mougeot}
\affiliation{CERN, 1211 Geneva, Switzerland}
\affiliation{CSNSM-IN2P3-CNRS, Universit\'{e} Paris-Sud, 91405 Orsay, France}

\author{L. Schweikhard}
\affiliation{Universit{\"a}t Greifswald, Greifswald, Germany}

\author{A. Welker}
\affiliation{CERN, 1211 Geneva, Switzerland}

\author{F. Wienholtz}
\affiliation{TU Darmstadt, Darmstadt, Germany}

\date{\today}

\begin{abstract}
Mass measurements of $^{49,50}$Sc, $^{70}$As, $^{73}$Br and $^{196}$Hg produced at CERN's radioactive beam facility ISOLDE are presented. The measurements were performed using the multi-reflection time-of-flight mass spectrometry and time-of-flight ion-cyclotron-resonance techniques with the ISOLTRAP mass spectrometer. The new results agree well with literature values, improving mass accuracy for all isotopes. 

\end{abstract}

\maketitle
\section{Introduction}

One of the fundamental properties of the atomic nucleus is its mass, which reflects nuclear structure through its binding energy \cite{2003Lunney}. There is an enormous amount of mass data collected over more than a hundred years of experimental investigation. The results of masses measured using different methods have been collected in the Atomic Mass Evaluation \cite{Huang2017}, the analysis of which supports the modern understanding of the evolution of the mass surface through the application of various mass filters along isotopic, isotonic or isobaric chains. Irregularities on the mass surface are revealed as nuclear-structure effects such as shell closures or deformation. Precise mass measurements facilitate finding these irregularities. Furthermore, new masses are essential to constrain the knowledge of astrophysical processes such as those responsible for the formation of elements heavier than iron. Nucleosynthesis and associated nuclear-mass models are heavily dependent on the accurate knowledge of atomic masses \cite{Pignatari,Str}.

Masses can be precisely measured using the revolution frequency of ions in a storage ring \cite{Franzke2008} or using time-of-flight ion-cyclotron resonance (TOF-ICR) \cite{Blaum2006}, employing Penning traps to determine ion's cyclotron frequency. 
A recently developed technique is multi-reflection time-of-flight mass spectrometry (MR-ToF MS) \cite{Wolf2013}, determining the flight time of ions traveling a path of about a kilometer folded into a table-top device. 
The latter two techniques are integral parts of ISOLTRAP's approach and have been used in this work to determine the masses of $^{49,50}$Sc, $^{70}$As, $^{73}$Br and $^{196}$Hg. 

\begin{figure*}
\includegraphics[width=\textwidth]{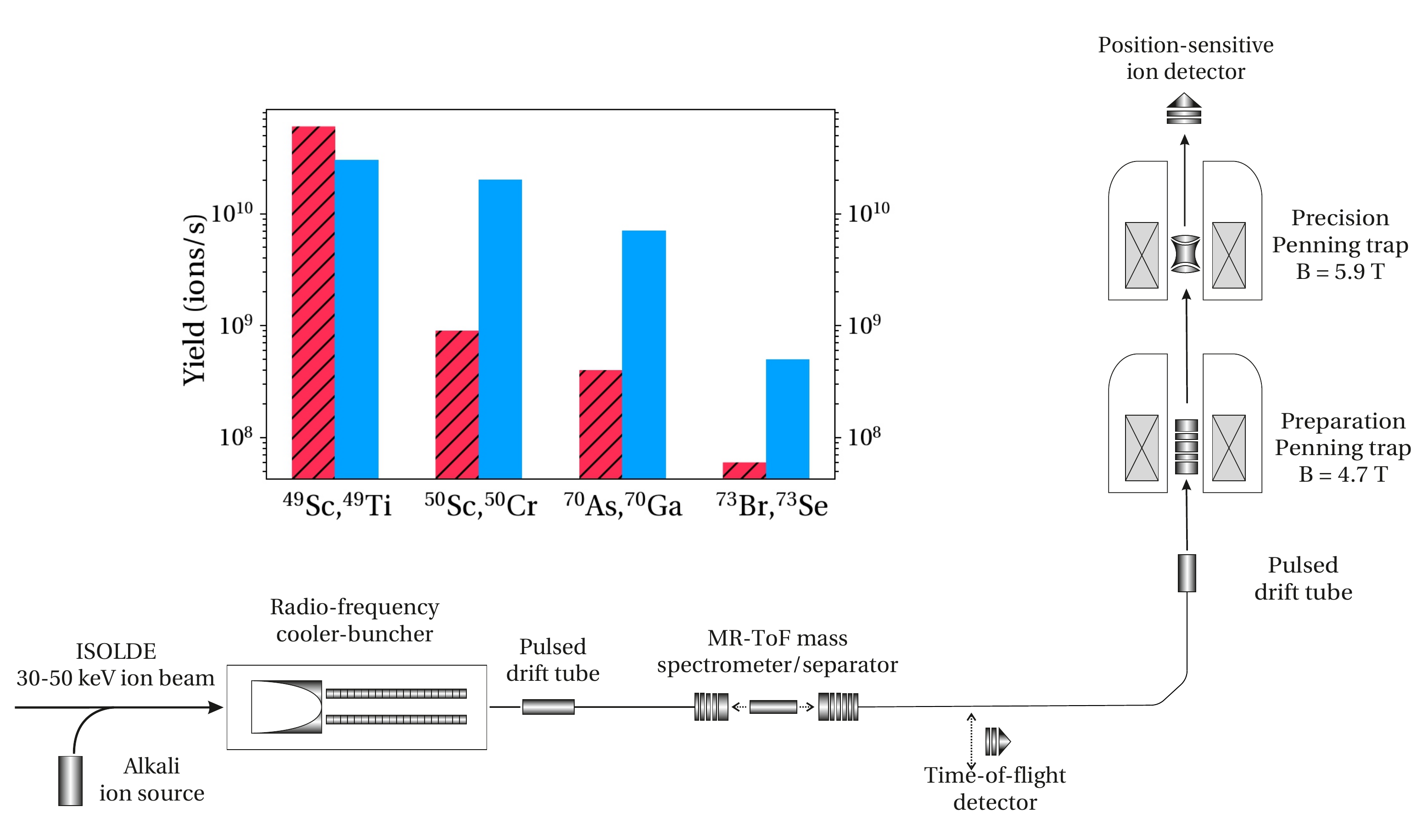}
\caption{\label{fig:1} Schematic view of the ISOLTRAP mass spectrometer with main components indicated (see text for details). Inset: yield estimation of $^{49}$Sc,$^{50}$Sc, $^{70}$As and $^{73}$Br (shown in red) and the main stable isobaric contaminant (in blue). The yield was determined using a time-of-flight detector placed downstream of the MR-ToF MS and corrected for the 0.25\% overall transport and detection efficiency.}
\end{figure*}

\section{Experiment}
\subsection{Isotope Production}

The radioactive isotopes of interest were produced at the ISOLDE facility at CERN \cite{Borge2018}, where a primary beam of 1.4-GeV protons impinged on the target with a typical thickness of several tens of g/cm$^{2}$ \cite{Hagebo1992}. The radioactive species then diffuse into an ion source, from which they are accelerated electrostatically, forming a beam that is transported through a dipole isotopic mass separator. The measurements were performed in three separate campaigns.
A tantalum-foil target was used to produce scandium isotopes that were ionized by using the highly element-selective ISOLDE Resonant Ionisation Laser Ion Source (RILIS)  \cite{Fedosseev2017, DayGoodacre2016}. 
The bromine and arsenic isotopes were produced using a zirconium-oxide powder target and ionized with a hot plasma ion source  
\cite{Kirchner1976,MartinezPalenzuela2018}. The mercury isotope was produced by an uranium carbide target, with a cold plasma ion source in VADIS mode. The ions of interest were selected by employing the General Purpose and High-Resolution separators of the ISOLDE facility.

\subsection{The ISOLTRAP Spectrometer}
A schematic representation of the ISOLTRAP high-precision mass spectrometer is shown in Fig. \ref{fig:1}. 
The apparatus is composed of four ion-trapping devices \cite{Mukherjee2008,Kreim2013}.
First, the quasi-continuous ISOLDE beam was injected into a linear radio-frequency quadrupole (RFQ) \cite{Herfurth2001}, 
where the beam was cooled through collisions with helium buffer gas. After a few milliseconds of beam cooling short ion bunches were extracted from the device, and their kinetic energy was adjusted to 3.2 keV using a pulsed drift cavity.
Afterward, the beam was injected into the MR-ToF device \cite{Wolf2013}, where the bunch was trapped using two electrostatic mirrors, between which the ion bunch reflected. The achieved ion-trajectory length was close to 1 km  (typical trapping time was $\sim$ 20 ms for $A \sim$ 85), causing the various species of different masses constituting the bunch to separate in time. By carefully adjusting the ejection time from the MR-ToF device, a purified ion ensemble could be delivered to the subsequent traps \cite{Wienholtz2017,Wolf2011}. Additionally, the MR-ToF MS can be used as a mass spectrometer on its own.

The resolving power of the MR-ToF device combined with its single-ion counting capability makes it particularly well-suited to study the composition of the ISOLDE beam \cite{Wolf2012}. The insert in Fig. \ref{fig:1} shows the yield of the isotopes measured in this work and of the most abundant stable isobaric contaminants measured using the ISOLTRAP MR-ToF MS. The yield of $^{49}$Sc was 10$^{10}$ ions per second, and the beam was contaminated by the isobar $^{49}$Ti. In the case of $^{50}$Sc the isobaric contaminants were $^{50}$Cr and $^{50}$Ti. It was still possible to perform the mass measurement, although the accuracy is affected by the high rate of stable isobaric contaminants. This contamination also hindered the precise mass measurements of scandium isotopes up to mass number $A=52$. The ratio of the yields of $^{73}$Se and $^{73}$Br was about 10. Figure~\ref{fig:2} (top) shows the TOF spectra of  $^{73}$Br isotope and its corresponding isobaric contaminants.  In the case of $^{70}$As the yield was about 10$^{8}$ ions per second, and the beam was contaminated mostly by $^{70}$Ga, $^{70}$Ge and $^{70}$Zn isobars.

The cylindrical preparation Penning trap \cite{RAIMBAULTHARTMANN1997378} received purified ion bunches from the MR-ToF mass separator. 
Here, the ion bunch can be further cooled and purified by using the well-established mass-selective resonant buffer-gas cooling \cite{Savard1991}. The preparation Penning trap wasn't used for purification in the discussed experiments. From the preparation Penning trap, the bunch was extracted and transferred to the precision Penning trap, where the mass measurements were performed. Penning-trap mass measurements are based on the so-called time-of-flight ion cyclotron resonance technique (TOF-ICR) \cite{Graff1980}, which determines the free cyclotron frequency $\nu_{c,x}$ of an ion inside the trap.

\subsection{Mass measurement techniques}
\begin{figure}[h!]
\includegraphics[width=\linewidth]{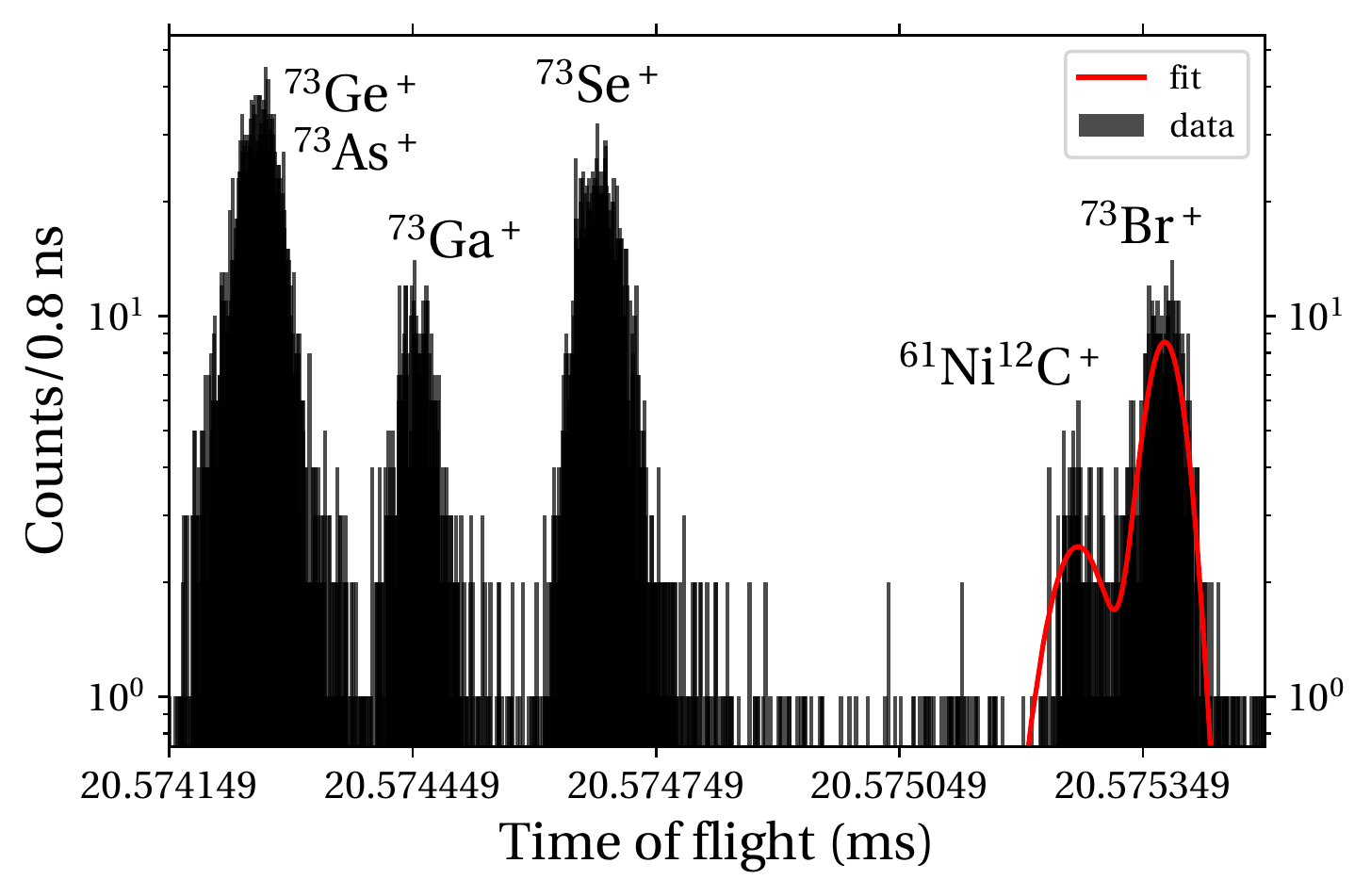}
\includegraphics[width=\linewidth]{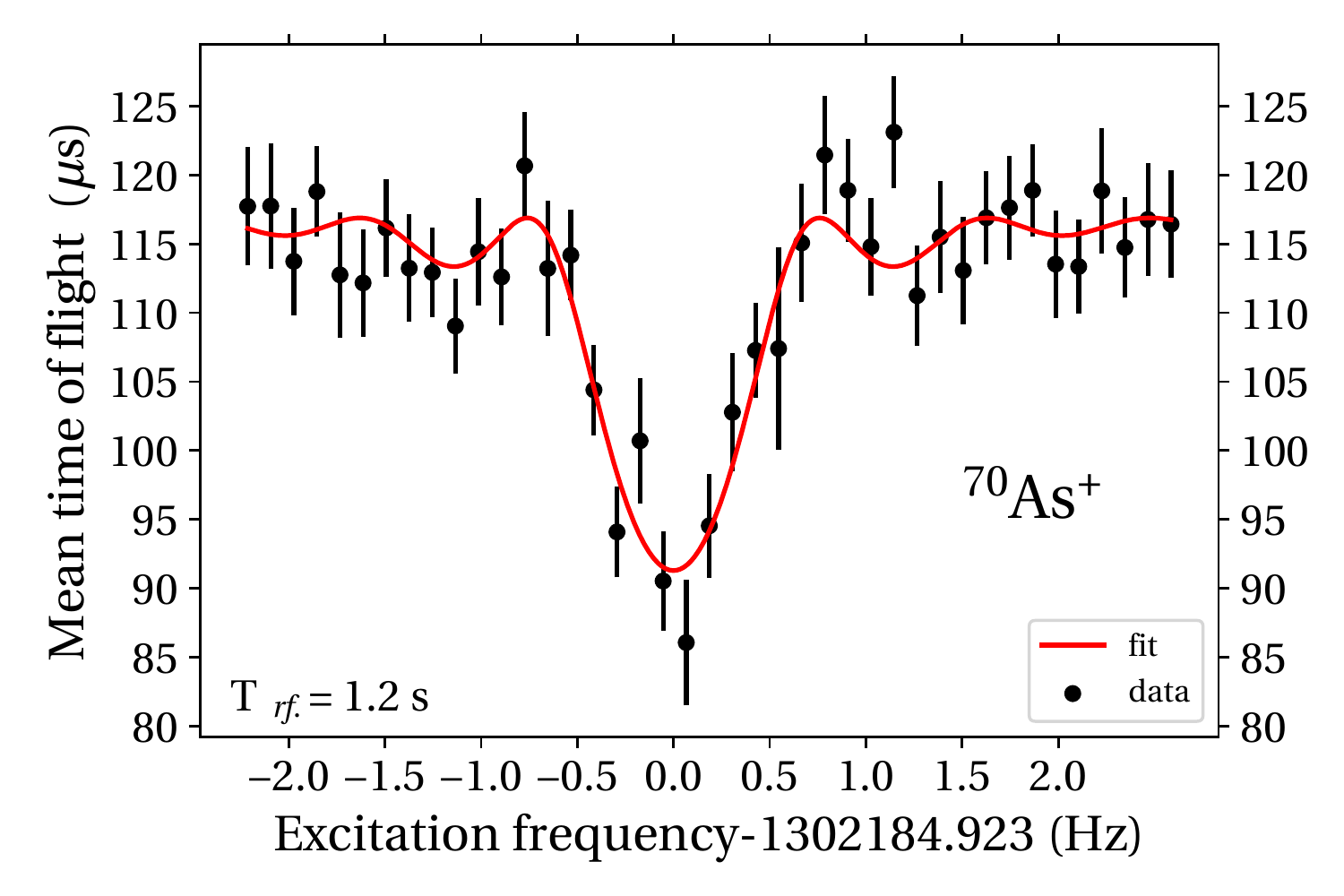}
\caption{\label{fig:2} Top: Time of flight of $^{73}$Br$^{+}$ and isobaric contaminants after 500 revolutions inside the MR-ToF MS. Bottom: a typical TOF-ICR resonance of $^{70}$As$^+$ using a single rf-excitation pulse of 1.2 seconds. The line represents a fit of the data points determined by the theoretical lineshape \cite{Koenig-IntJMassSpectrom.142.95}}.
\end{figure}

Due to its excellent resolving power (often exceeding 100,000) the MR-ToF device can also be used as a mass spectrometer \cite{Wienholtz-Nature.498.346}. In this case the isobars can be used as references (provided their masses are known well enough).
Figure \ref{fig:2} (top) shows the measured time-of-flight (TOF) spectrum of the $^{73}$Br$^{+}$ isotope and its isobaric contaminants. The mass of an ion species can be calculated according to:
\begin{equation}
    t=\alpha\sqrt{m/q}+\beta , 
\end{equation} where $m$ is the ion's mass, $q$ its charge state and $\alpha$ and $\beta$ are calibration parameters. These parameters have to be evaluated by measuring the TOFs $t_{1,2}$ of two reference species with well-known masses $m_{1,2}$. The mass over charge ratio of the ion of interest can then be determined by using \cite{Wienholtz-Nature.498.346}:
\begin{equation}
 \sqrt{m/q}= C_{TOF}\Delta_{Ref}+\frac{1}{2}\Sigma_{Ref},
\end{equation} where $C_{TOF}=\frac{2t-t_1-t_2}{2(t_1-t_2)}$ depends only on the TOF of the ion of interest and the TOFs of the reference ions, $\Delta_{Ref}=\sqrt{m_1}-\sqrt{m_2}$ and  $\Sigma_{Ref}=\sqrt{m_1}+\sqrt{m_2}$ depend on the reference-ion masses. Usually one employs well-known calibrants such as $^{39}$K$^+$, $^{85}$Rb$^+$ or $^{133}$Cs$^+$, provided by an offline ion source. However, to minimize  systematic uncertainties coming from the different trajectories of offline and online beams, one typically uses at least one of the online isobaric contaminants as calibrants.

The other available mass measurement technique is TOF-ICR. The mass is determined via the measurement of the cyclotron frequency $\nu_c=\frac{qB}{2\pi m}$ of ions with charge $q$ stored in the homogeneous magnetic field $B$. Ions are captured and stored in the center of the Penning trap. The ion dynamics is composed of three eigenmotions with three characteristic frequencies: the axial $\nu_z$, magnetron $\nu_-$ and modified cyclotron $\nu_+$ \cite{Blaum2006}. By applying a quadrupole excitation pulse to the segmented trap electrodes for time duration $T_{rf}$ \cite{Koenig-IntJMassSpectrom.142.95,George2007}, one can couple both radial motions $\nu_-$ and $\nu_+$. The ions are extracted from the trap through an electrical drift tube to a micro-channel plate (MCP) detector. Due to the magnetic field gradient from the high-field region inside the trap to nearly field-free region at the MCP, the energy of the radial motion is converted into an axial one leading to an additional velocity boost of the ions. If the frequency of the quadrupole excitation matches the cyclotron frequency of the ions, a resonant conversion is fulfilled, and the ions travel the distance to the MCP detector in the shortest time. Hence, the time-of-flight method can be applied to determine $\nu_c$. The magnetic-field induction $B$ has to be stable and known with high precision. For this purpose a superconducting magnet is used. Calibration of the magnetic field is done by using well-known reference masses provided from the offline ion source. The cyclotron frequency $\nu_{c,ref}$ of one of the $^{39}$K$^+$, $^{85}$Rb$^+$ or $^{133}$Cs$^+$ reference isotopes is measured before and after measuring $\nu_c$ of the ions of interest to take into account any possible variations of the magnetic field. If the ion species are singly charged $q=+e$, the atomic mass is given by:
\begin{equation}
 m= \frac{\nu_{c,ref}}{\nu_c}(m_{ref}-m_e)+m_e
\end{equation}
where $m_e$ is the electron mass and $m_{ref}$ is the mass of the reference ion. To account for a possible change in future measurements of the employed reference masses, the mass of the isotope of interest is reported as a measured primary frequency ratio $r_{icr}=\frac{\nu_{c,ref}}{\nu_c}$.

\begin{figure*}
        \begin{subfigure}[b]{0.475\textwidth}
            \centering
            \includegraphics[width=\textwidth]{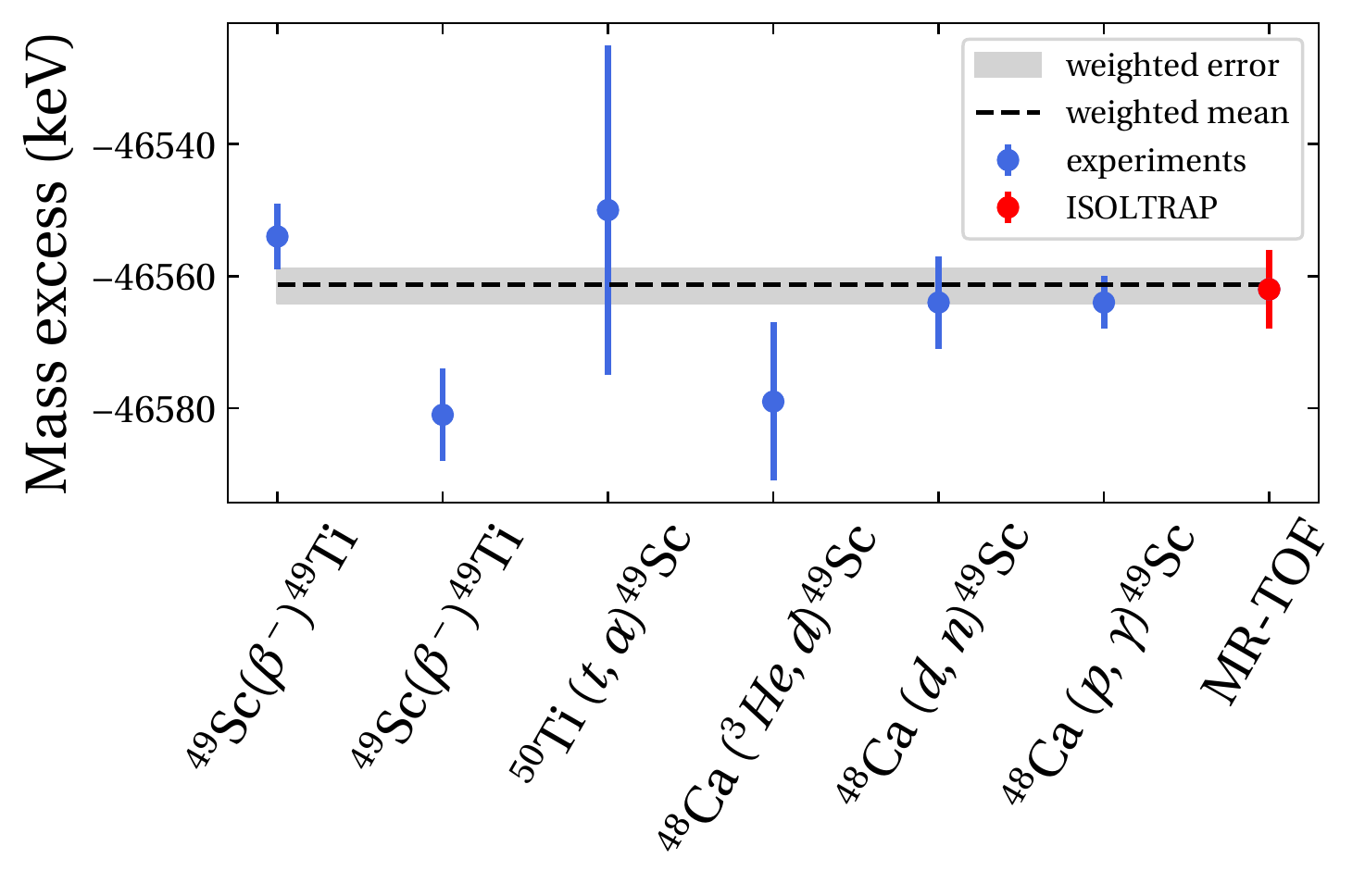}
            \caption[Network2]%
            {{\small The experimental mass excess values of $^{49}$Sc obtained from \cite{Rezanka1961,Flothmann1969,Erskine1966, Williams1966,Grandy1968,Vingiani}. For better visibility values from \cite{Kelley1956,Martin1956} are excluded from the plot since they have uncertainties of 50 and 100 keV.}}    
            \label{fig:4_1}
        \end{subfigure}
        \hfill
        \begin{subfigure}[b]{0.475\textwidth}  
            \centering 
            \includegraphics[width=\textwidth]{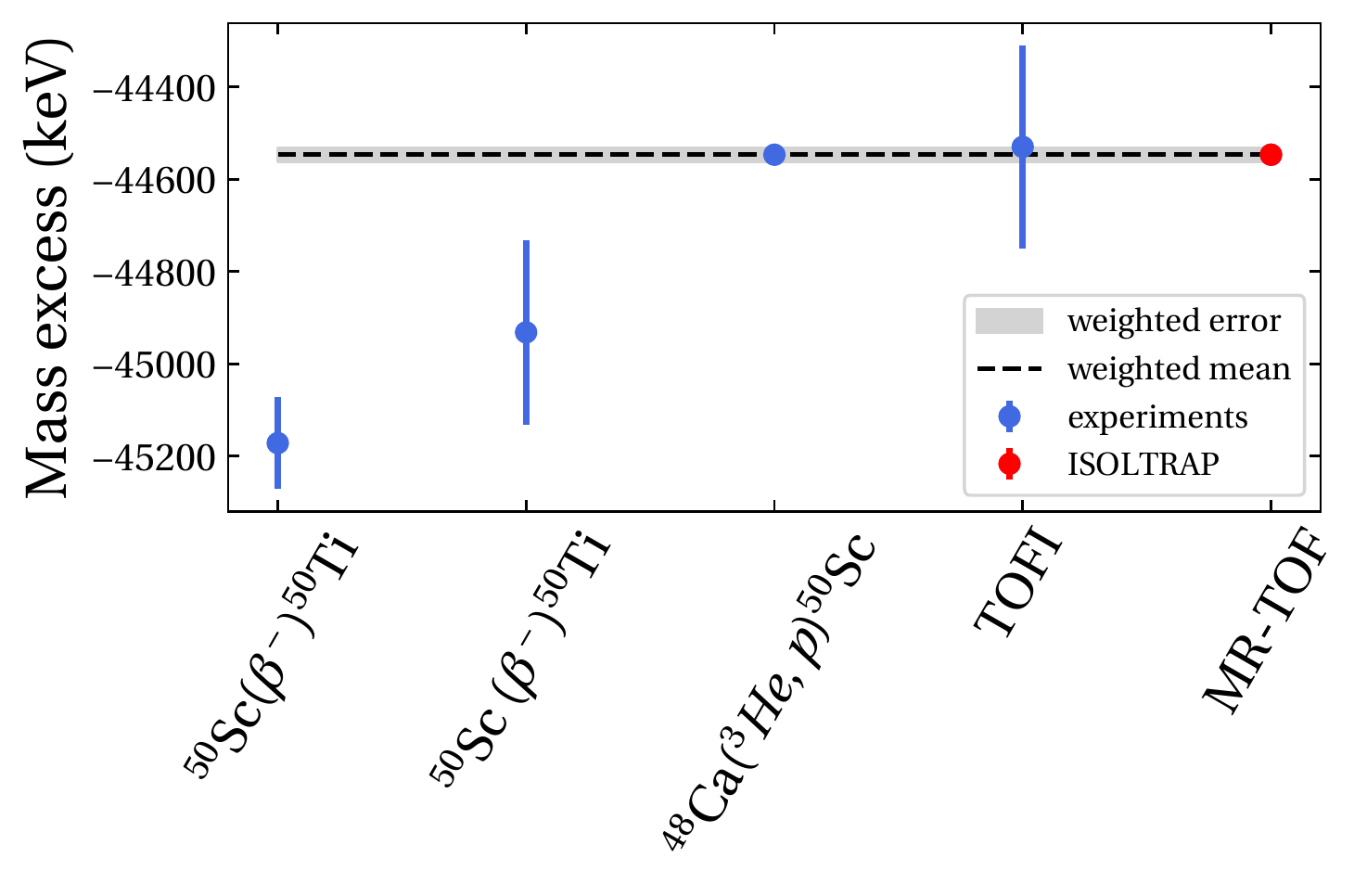}
            \caption[]%
            {{\small The experimental mass excess values of $^{50}$Sc obtained from \cite{Ward,Chilosi,Ohnuma,Bai}.}}
            \label{fig:4_2}
        \end{subfigure}
        \vskip\baselineskip
        \begin{subfigure}[b]{0.475\textwidth}   
            \centering 
            \includegraphics[width=\textwidth]{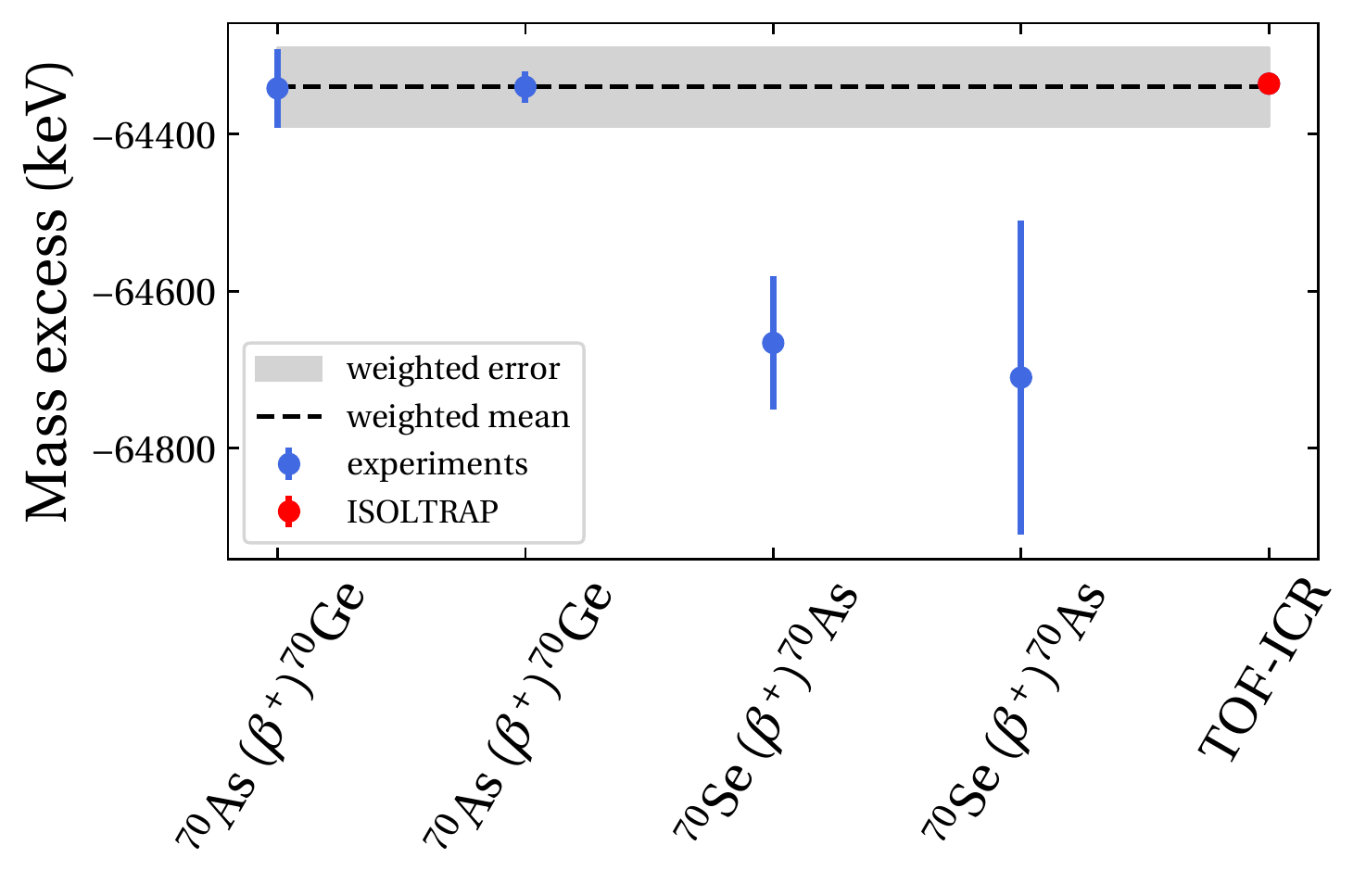}
            \caption[]%
            {{\small The experimental mass excess values of $^{70}$As obtained from \cite{Born1963,SHEN2017, Labreque1975,Tomlin2001}.}}    
            \label{fig:4_3}
        \end{subfigure}
        \quad
        \begin{subfigure}[b]{0.475\textwidth}   
            \centering 
            \includegraphics[width=\textwidth]{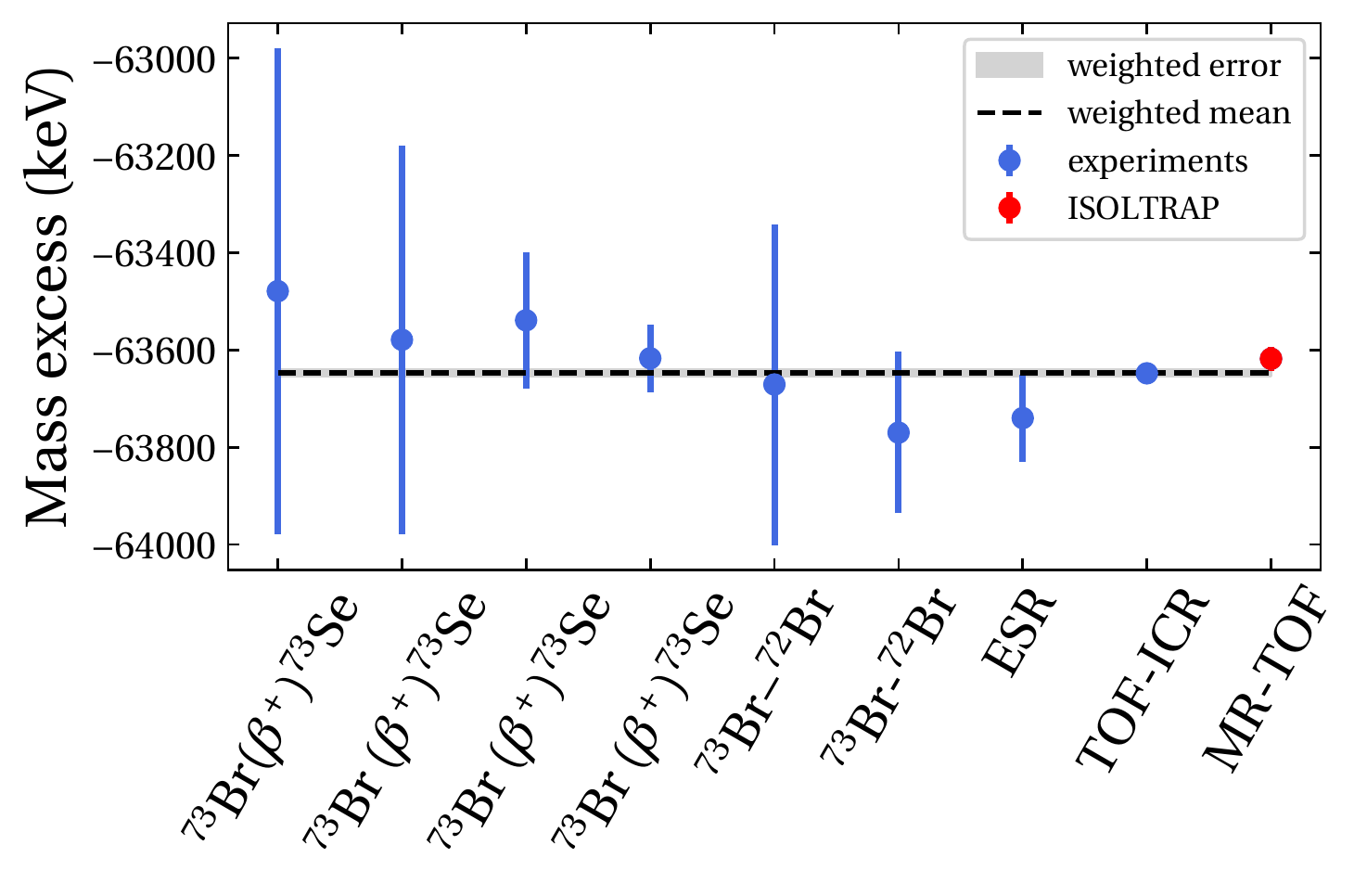}
            \caption[]%
            {{\small The experimental mass excess values of $^{73}$Br obtained from \cite{Tomlin2001,Roeckl1974,HEIGUCHI1987484,Murray, Hausmann2001, Herfurth2011,Sharma1989,Sharma1991}.}}    
            \label{fig:4_4}
        \end{subfigure}
        \caption[  ]
        {\small The dashed line and the gray area represent the weighted mass excess value and its uncertainty as calculated by AME16 based on the measurements shown as dots. The new mass excess values (red dot) are not included in the evaluation. The contribution of the new mass excess values to the updated AME results can be deduced from the last column of Table 1.} 
        \label{fig:4}
    \end{figure*}

\section{Results}

 \begin{table*}
\caption{\label{tab:table1} Summary of the $r_{icr}$ and ${C_{TOF}}$ ratios, as well as mass excess values of $^{49,50}$Sc, $^{70}$As, $^{73}$Br and $^{196}$Hg. Reference masses taken from AME2016 \cite{AME} and $T_{1/2}$ values from NUBASE16 \cite{Nubase}. The last column are the significances of the ISOLTRAP mass measurements in the re-evaluated mass excess values.}
\begin{ruledtabular}
\begin{tabular}{ccccccccc}
&\multicolumn{2}{c}{}&\multicolumn{2}{c}{}&\multicolumn{1}{c}{Mass excess (keV)}&\multicolumn{2}{c}{}\\
\cline{5-7} 
Isotope&$T_{1/2}$&References&$r_{icr}$, $C_{TOF}$&This work&AME16&re-evaluated&Sign.\\ 
\hline
 
 $^{49}$Sc& 57.2 min&$^{49}$Ti, $^{39}$K& $C_{TOF}$=0.500203604(622)& -46562.3(6.1)& -46561.3(2.7)&-46561.6(2.5)&16 \%\\
 $^{50}$Sc& 102.5 sec&$^{50}$Ti, $^{39}$K&$C_{TOF}$=0.500633802(837) & -44546.5(9.1)& -44547(15)& -44546.8(7.8)&73 \%\\ 
 $^{70}$As& 52.6 min&$^{85}$Rb& $r_{icr}$=0.8235704146(180)& -64334.0(1.4)& -64340(50)& -64334.0 (1.4)&100 \%\\
 $^{73}$Br& 3.4 min&$^{73}$Se, $^{85}$Rb&$C_{TOF}$=0.499572415(1620)& -63635.3(19)& -63647(7)&-63646.4(6.9)&9 \%\\
  $^{196}$Hg& stable&$^{39}$K& $r_{icr}$=2.5147441131(1500)&-31818.8(10.9)&-31825.9(2.9)&-31824.3(2.6)&23 \%\\
\end{tabular}
\end{ruledtabular}
\end{table*} 

Table \ref{tab:table1} summarizes the mass excess values for the measured nuclides and Fig. \ref{fig:4} represents the comparison of the new mass excess values with the ones provided in the literature, taken from the Atomic Mass Evolution 2016 \cite{AME}. 

\subsection{$^{49}$Sc}
The first mass determination of $^{49}$Sc was derived using the $Q$-values of $^{49}$Ca($\beta^-$)$^{49}$Sc \cite{Kelley1956,Martin1956} and $^{49}$Sc($\beta^-$)$^{49}$Ti decays \cite{Rezanka1961,Flothmann1969}. A few years later, this nucleus was used in several nuclear reaction experiments giving additional $Q$-values \cite{Erskine1966, Williams1966,Grandy1968,Vingiani}. The number of $\beta^-$ decay and reaction experiments done with this nuclide constrained the mass excess value to -46561.3(2.7) keV. The new mass excess value obtained by our MR-ToF MS measuremnts agrees well with AME16, deviating by merely 1 keV. Moreover, the significance of the new direct mass excess value in the re-evaluated AME is 16 \%. The new MR-ToF MS result is the first direct mass measurement of $^{49}$Sc.

\subsection{$^{50}$Sc}
 The $^{50}$Sc($\beta^-$)$^{50}$Ti decay \cite{Ward,Chilosi} and the $^{48}$Ca(He$^{3}$,p)$^{50}$Sc reaction \cite{Ohnuma} were the first experiments to provide a mass excess value of $^{50}$Sc. The first direct mass measurement was performed using the TOFI spectrometer, in 1998 \cite{Bai}. All together they defined the mass excess value to -44547(15) keV. The uncertainty was mainly determined by the result from the $^{48}$Ca(He$^{3}$,p)$^{50}$Sc reaction, which is 15 times more precise than the uncertainty obtained by TOFI. The new mass excess value measured by the MR-ToF MS technique with the uncertainty of 9.1 keV agrees perfectly with the AME16 values and now contributes 73\% to the re-evaluated mass excess value.
 
\subsection{$^{70}$As}
The mass of $^{70}$As was known from $Q$-values of $^{70}$As($\beta^+$)$^{70}$Ge \cite{Born1963,SHEN2017} and $^{70}$Se($\beta^+$)$^{70}$As decays \cite{Labreque1975, Tomlin2001}. The mass excess value was determined more precisely from the $^{70}$As($\beta^+$)$^{70}$Ge decay, yielding -64340(50) keV. $^{70}$Ge is the stable isotope with well-known mass. Therefore, the mass excess of $^{70}$As mostly determined by the $^{70}$As($\beta^+$)$^{70}$Ge decay. The new mass excess value measured by  TOF-ICR technique is 36 times more precise.Therefore, the re-evaluated mass excess value is now entirely determined through this ISOLTRAP measurement.

\subsection{$^{73}$Br}
 There are seven previous measurements involving $^{73}$Br: four beta decay studies  \cite{Tomlin2001,Roeckl1974,HEIGUCHI1987484,Murray}, one ESR storage ring experiment\cite{Hausmann2001}, a TOF-ICR value measured by ISOLTRAP \cite{Herfurth2011} and two values obtained via deflection-voltage ratios of $^{73}$Br and $^{72}$Br using a magnetic dipole separator at the Chalk River TASCC facility \cite{Sharma1989, Sharma1991}. The mass excess value obtained from those experiments is -63647(7) keV and is mainly determined by the TOF-ICR result from ISOLTRAP reported in 2011. The new mass excess value obtained by MR-ToF MS agrees with the literature value within the error bars and contributes 9\% to the re-evaluated mass excess.

\subsection{$^{196}$Hg}
The mass of $^{196}$Hg was mostly determined from deflection-voltage ratios of $^{196}$Hg and $^{198}$Hg$^{35}$Cl molecule \cite{Kozier80}. Together With $^{196}$Au($\beta^-$)$^{196}$Hg decay \cite{Waptsra62} it yielded the total mass excess of -31825.9(2.9) keV. The mass of $^{196}$Hg was determined by ISOLTRAP but the unexpected difference with the weighted literature value led to rejection of the mass measurement \cite{Schwarz2001}. The new mass value measured by TOF-ICR technique is in very good agreement with all literature values and contributes 23\% to the new re-evaluated value.
\begin{figure}[h!]
\includegraphics[width=\linewidth]{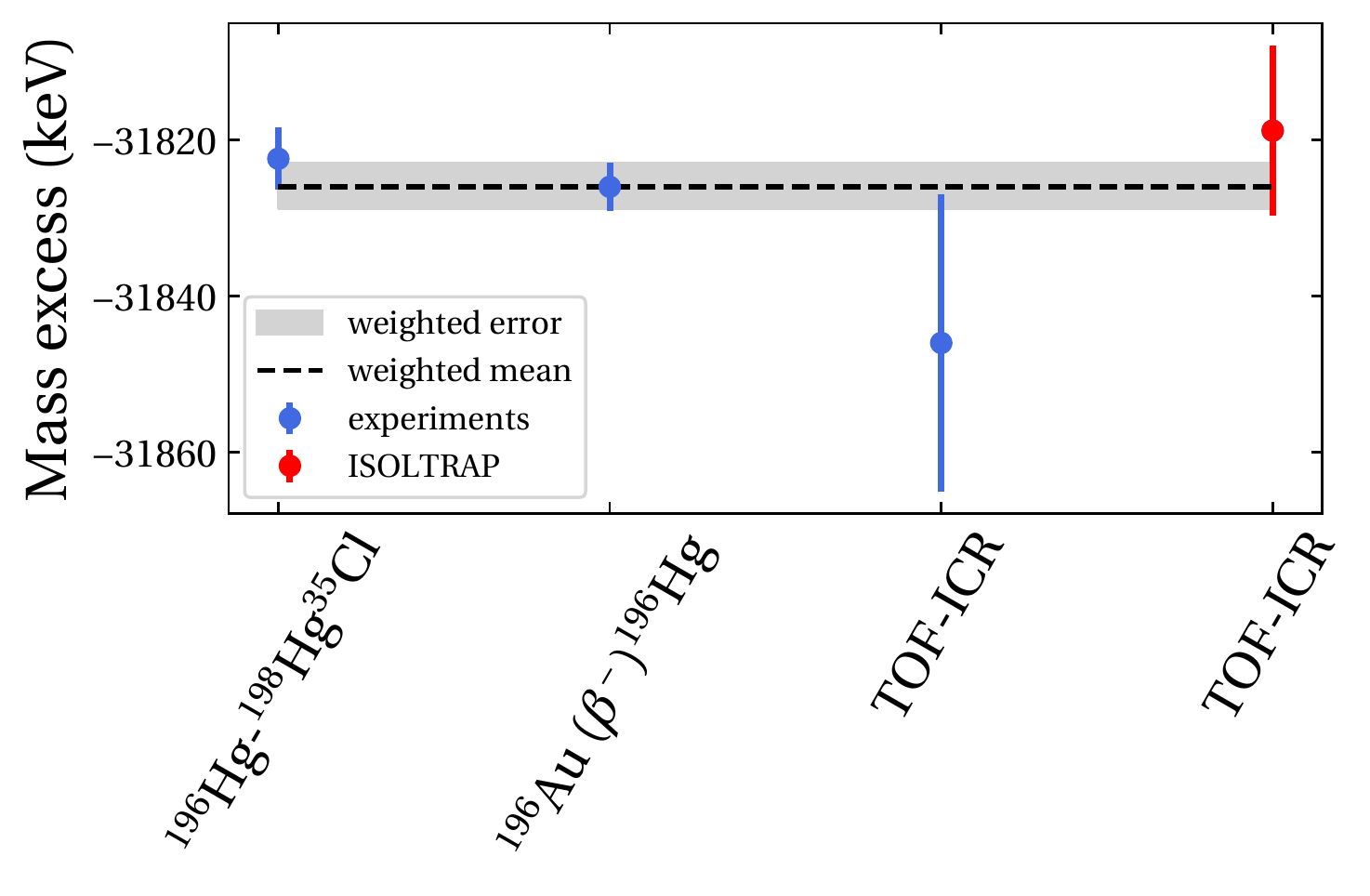}
\caption{\label{fig:5} The experimental mass excess values of $^{196}$Hg obtained from \cite{Kozier80,Waptsra62,Schwarz2001}.}
\end{figure}

\begin{figure}[h!]
\includegraphics[width=\linewidth]{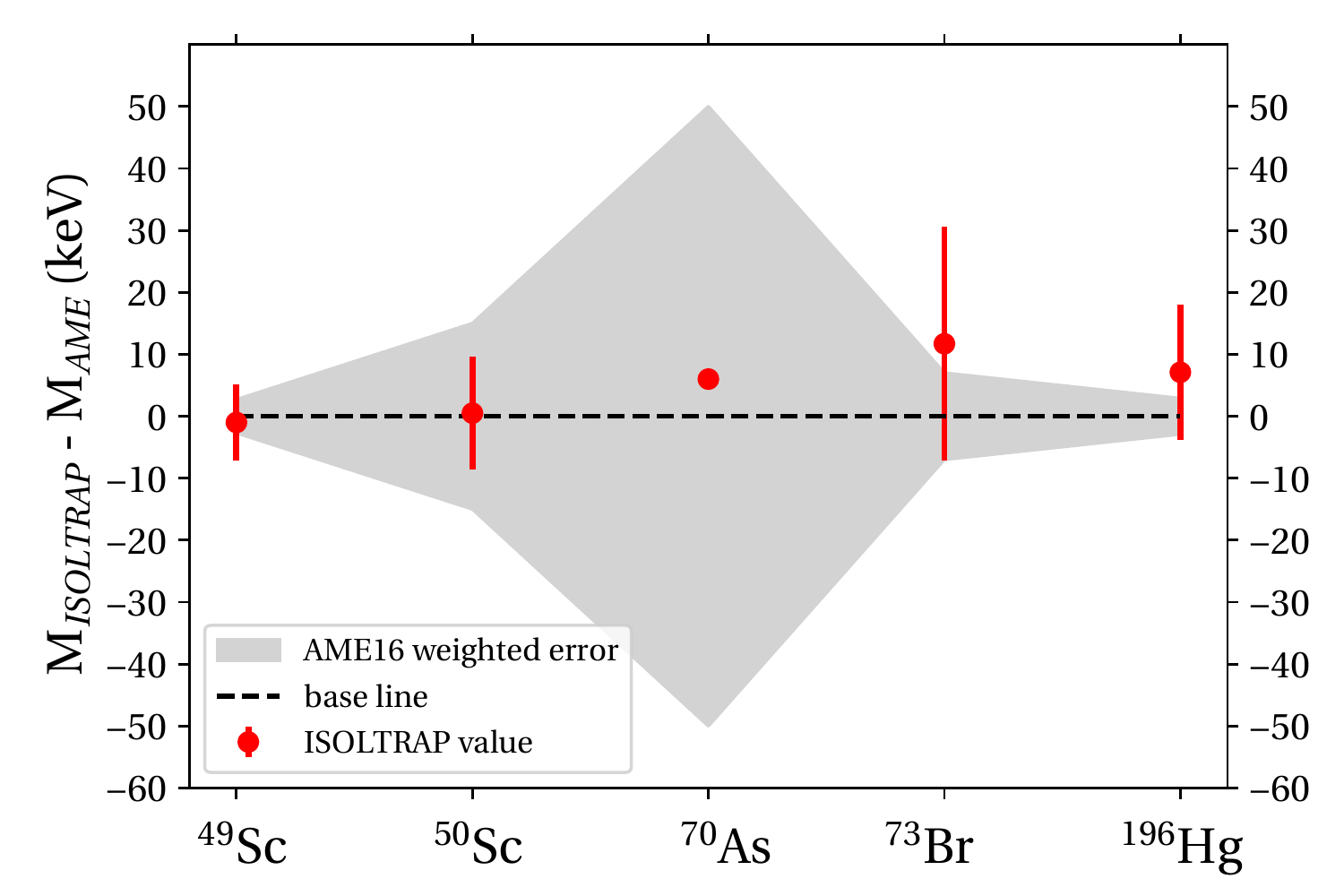}
\caption{\label{fig:6} Differences in the mass excess values between our results and AME2016. 
The gray shaded background represents the AME16 uncertainty.}
\end{figure}

The results are summarized in Fig.~\ref{fig:6}.  Overall, our new experimental values agree with the literature values within the reported errors.
The uncertainties were reduced by factors 1.6 and 35 for $^{50}$Sc and $^{70}$As nuclei, respectively. These two isotopes gave the biggest impact on the re-evaluated values in the updated AME16. All isotopes listed above can be used as online references in the future experiments of ISOLTRAP.
\section{Summary}
The masses of the short-lived $^{49,50}$Sc, $^{70}$As and $^{73}$Br and stable $^{196}$Hg nuclides have been investigated by the versatile mass spectrometer ISOLTRAP with uncertainties on the level of a few keV. The mass precision has been improved in all cases. No deviations from literature values were observed. The new results significantly contribute to the corresponding re-adjusted mass values in the global atomic mass evaluation. The agreement of the new data obtained with the MR-ToF MS confirms the reliability of this complementary mass measurement technique.

\section{Acknowledgements}
We thank the ISOLDE technical group and the ISOLDE Collaboration for their support. We acknowledge the support of the Max Planck Society, the French Institut National de Physique Nucleaire et de Physique des Particules (IN2P3), the European Research Council (ERC) through the European Union's Horizon 2020 research and innovation program (grant agreement No 682841 ''ASTRUm''), and the Bundesministerium f{\"u}r Bildung und Forschung (grants No. 05P15ODCIA and No. 05P15HGCIA). Jonas Karthein acknowledges the support of a Wolfgang Gentner Ph.D. scholarship of the BMBF (05E12CHA).

\bibliography{My_Collection}
\bibliographystyle{unsrtnat}

\appendix
\section{Appendix}
\begin{figure}[h!]
\includegraphics[width=\linewidth]{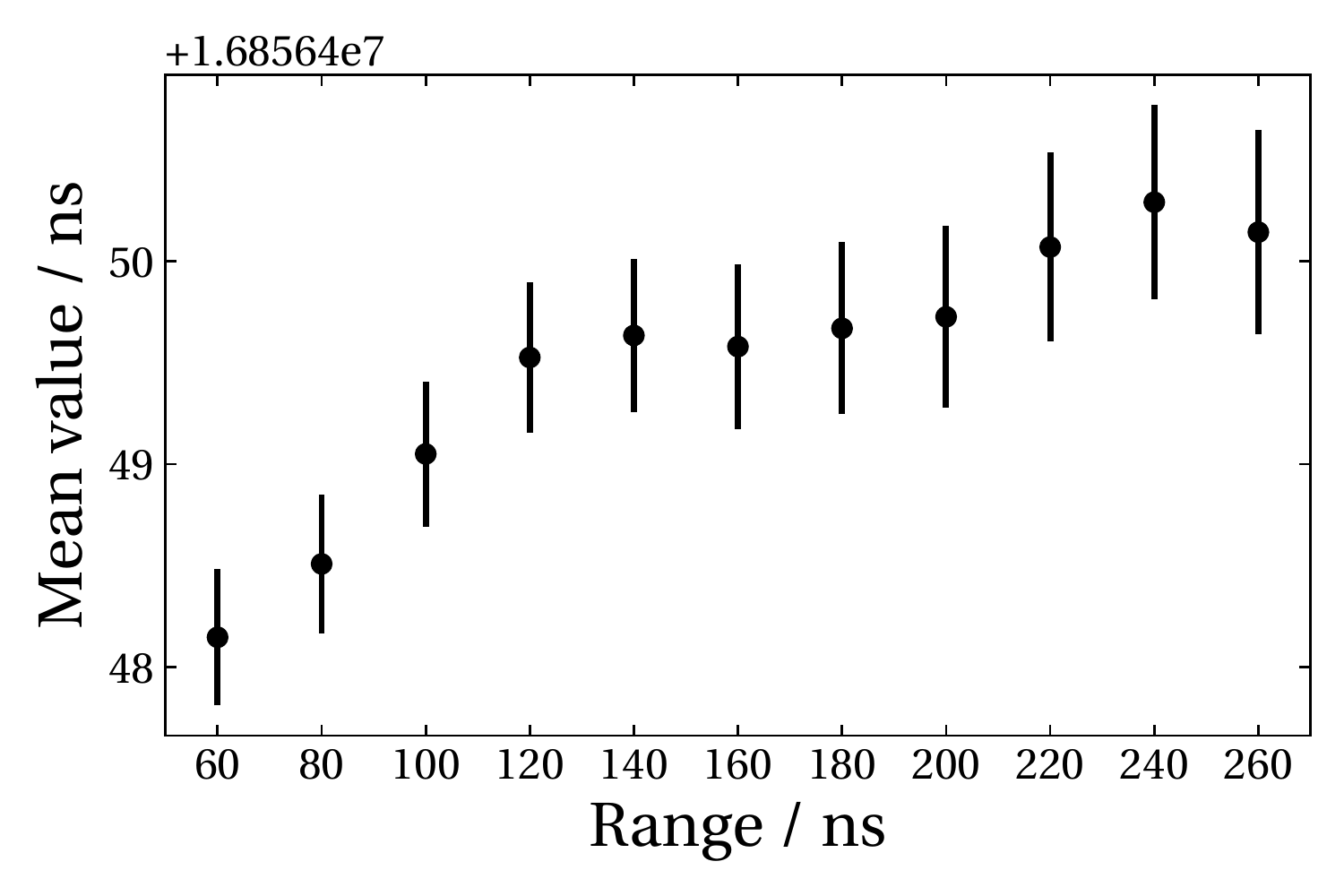}
\includegraphics[width=\linewidth]{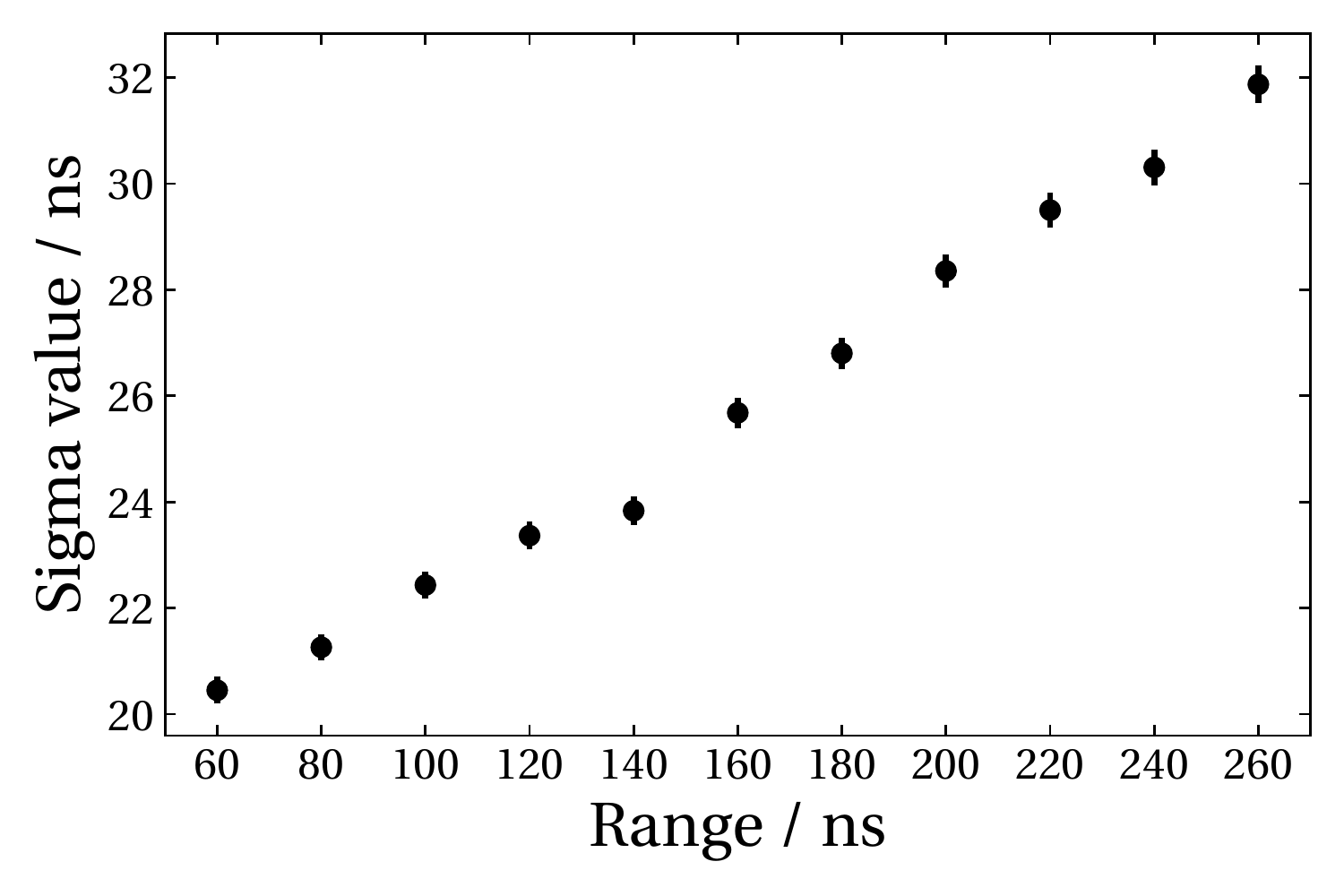}
\caption{\label{fig:3} Top: the evolution of the mean value of the Gaussian distribution of $^{49}$Sc isotope. Bottom: the evolution of the sigma value of the Gaussian distribution for $^{49}$Sc}.
\end{figure}

The data analysis of the MR-ToF spectra was based on the binned maximum-likelihood estimation. Various TOF selection windows were selected while the probability density functions used for the parameter estimation were Gaussian, Double Gaussian and Exponential Gaussian Hybrid (EGH) \cite{LAN20011}. It resulted in the weighted $\overline{C_{TOF}}$ for the ion of interest and the corresponding statistical uncertainty $\sigma_{\overline{C_{TOF}},stat}$. The difference in $\overline{C_{TOF}}$ determined by two different fit functions is considered to be a part of the systematic error and has been added in quadrature as $\sigma_{C_{TOF},pdf}$ to the total uncertainty. This error accounts for the imperfection of the peak shape. By changing the fitting range, the second part of the systematic error coming from the imperfection of the peak shape $\sigma_{C_{TOF},window}$ has been obtained. The evolution of Gaussian parameters is shown in Figure \ref{fig:3}. The window range was chosen according to these evaluations for every data set individually. The presents of the the plateau was an indicator of stable output fit parameters but was not observed for every peak.
Although both checks (varying the fit function and varying the fit window) have a contribution to the error due to the peak asymmetry, a conservative approach was taken in this work and the two resulting systematic uncertainties were treated as independent. Finally, the total uncertainty can be expressed as
 $\sigma_{C_{TOF},total}=\sqrt{(\sigma_{\overline{C_{TOF}},stat})^2+(\sigma_{\overline{C_{TOF}},pdf})^2+(\sigma_{\overline{C_{TOF}},window})^2}$. The uncertainties are described below for each nuclide, and the final results are listed in Table 1.
 
\subsection{$^{49}$Sc}
Ions of $^{49}$Sc were stored in the MR-ToF device for 500 revolutions. Approximately 14000 $^{49}$Sc ions were recorded in three data sets. The TOF spectra of $^{49}$Sc$^+$ were fitted by Gaussian and EGH fit functions. The $^{49}$Ti ions were used as an online reference, and $^{39}$K$^+$ was chosen as an offline reference. Online and offline references were likewise fitted by Gaussian and EGH fit functions. The Gaussian fit function was used for computing the $\overline{C_{TOF}}$. The weighted statistical uncertainty was $2.6\times10^{-7}$. Although, only three data spectra were taken the $\chi^{2}$ test was performed ($\chi^{2}$=1.42). The difference between scandium $\overline{C_{TOF}}$ values obtained from Gaussian and EGH fits was $5.5\times10^{-7}$. This was added in quadrature to the total uncertainty as $\sigma_{C_{TOF},pdf}$. The optimal fitting window was chosen to be 120 ns. The obtained results for 120 ns and 100 ns fitting windows differed by $1.34\times10^{-7}$. For fitting windows of 120 ns and 140 ns, the corresponding difference was $1.40\times10^{-8}$. Both values were added in quadrature to the total uncertainty as $\sigma_{C_{TOF},window}$. The weighted $\overline{C_{TOF}}=0.500203604$ with total uncertainty of $6.22\times10^{-7}$.

\subsection{$^{50}$Sc} 
Ions of $^{50}$Sc were stored in the MR-ToF MS for 500 revolutions. Altogether there were about 3600 ions of $^{50}$Sc in four data sets. The same analysis as for $^{49}$Sc$^+$ was made. The online reference was $^{50}$Ti$^+$, and the offline reference was $^{39}$K$^+$. The Gaussian fit function was used for computing $\overline{C_{TOF}}$. This led to the weighted $\overline{C_{TOF}}$=0.500633802 with statistical uncertainty $2.34\times10^{-7}$. The $\chi^{2}$ was found to be 8.2 and the statistical uncertainty was normalised to $6.73\times10^{-7}$. The difference due to the choice of PDF gave the systematic uncertainty of $4.01\times10^{-7}$. The systematic uncertainty from varying the fitting window amounted to $2.16\times10^{-7}$ and $1.98\times10^{-7}$ for smaller and larger windows, respectively. The weighted $\overline{C_{TOF}}=0.500633802$ with the total uncertainty of $8.37\times10^{-7}$.

\subsection{$^{70}$As}
The mass measurement of $^{70}$As$^+$ was done with the Penning trap by applying the TOF-ICR technique. A single-excitation-pulse TOF-ICR technique was used.
Figure \ref{fig:2} (bottom) shows the measured TOF spectrum of $^{70}$As$^+$.
Two TOF-ICR spectra were taken for two excitation times of 0.1 and 1.2 seconds. The reference spectrum of $^{85}$Rb ions was taken every time before and after the mass measurement to account for changes in the magnetic field. The total statistical uncertainty was $1.6\times10^{-8}$. The correction for the difference in the mass between the ion of interest and the reference ion added an error of $4.3\times10^{-9}$. To avoid ion-ion interaction shifts the number of ions detected per ejection was fixed to $\leq$ 5. Also, the residual systematic uncertainty $8\times10^{-9}$ \cite{Kellerbauer} of the ISOLTRAP mass spectrometer was added in quadrature. The frequency ratio is $r_{icr}$=0.8235704146 with the total uncertainty $1.8\times10^{-8}$

\subsection{$^{73}$Br}
The $^{73}$Br ions were stored in the MR-ToF MS for 250, 500 and 1000 revolutions. There were three data sets recorded with a total of about 4500 ions. The $^{61}$Ni$^{12}$C molecule \cite{shim} contaminated the spectra (see Fig. \ref{fig:2} top panel). The resolving power $R=\frac{m}{\Delta{m}}=\frac{t}{2\Delta{t}}$ was 123000, while the minimum resolving power required to resolve these ion species is 118000 as calculated from AME2016. Thus, it was impossible to completely resolve the $^{73}$Br$^+$ and $^{61}$Ni$^{12}$C$^+$ peaks. Therefore, a two-Gaussian fit function was used for computing the $\overline{C_{TOF}}$. To employ the EGH and Gaussian fit functions, the fitting window had to be reduced to 80 ns otherwise the bromine isotope and the $^{61}$Ni$^{12}$C molecule recognized as a single peak. The $\chi^{2}$ was 0.48 and no normalization was applied. The difference between the fit results with Gaussian and two-Gaussian fit functions was $2.17\times10^{-8}$ and the corresponding difference between EGH and two-Gaussian functions was $2.63\times10^{-7}$. The systematic errors $1.23\times10^{-6}$ and $1.0\times10^{-6}$ were obtained as above by varying the fitting window. The optimum window was 80 ns. This was the main source of systematic uncertainty. The final uncertainty of $1.63\times10^{-6}$ was determined.

\subsection{$^{196}$Hg}
The mass measurement of $^{196}$Hg$^{2+}$ was done with the Penning trap by applying the TOF-ICR technique with a single-excitation-pulse.
Four TOF-ICR spectra were taken. One spectrum with 0.4 second and tree others with 0.6 seconds of excitation time. The reference spectrum of $^{39}$K ions was taken every time before and after the mass measurement. The total statistical uncertainty was $5.4\times10^{-8}$. The dominant part in the uncertainty was the correction for the mass difference between $^{196}$Hg$^{2+}$ and reference $^{39}$K$^+$. It resulted $1.4\times10^{-7}$ of the mass dependent systematic uncertainty. Also, the residual systematic uncertainty $8\times10^{-9}$ \cite{Kellerbauer} of the ISOLTRAP mass spectrometer was added in quadrature. The number of ions detected per ejection was fixed to $\leq$ 5. Finally, the frequency ratio is $r_{icr}$=2.5147441131 with the total uncertainty $1.5\times10^{-7}$

\end{document}